\documentstyle[twocolumn,prb,aps,epsfig]{revtex}
\def\cuge{CuGeO$_3$\ }

\def\Jeff{$J_{\text{eff}}$\ }
\def\sus{susceptibility\ }
\begin{document}
\draft
\title{Thermodynamical Properties of a Spin-$\frac{1}{2}$ Heisenberg Chain 
Coupled to Phonons}

\vspace{1cm}

\author{Rainer W. K\"uhne}
\address{Fachbereich Physik, Universit\"at Wuppertal, 42097 Wuppertal, Germany}
\author{Ute L\"ow}
\address{Institut f\"ur Theoretische Physik, Universit\"at Dortmund, 44221 
Dortmund, Germany}
\maketitle

\vspace{1cm}

\begin{abstract}
We performed a finite-temperature quantum Monte Carlo simulation of the 
one-dimensional spin-$\frac{1}{2}$ Heisenberg model with nearest-neighbor 
interaction coupled to Einstein phonons. 
Our method allows to treat easily up to 100 phonons per site and 
the results presented are practically free from truncation errors. 
We studied in detail the magnetic susceptibility, the specific heat, the 
phonon occupation, the dimerization, and the spin-correlation function 
for various spin-phonon couplings and phonon frequencies. 
In particular we give evidence for the transition from a gapless 
to a massive phase by studying the finite-size 
behavior of the susceptibility. We also show that the dimerization is 
proportional to $g^2/\Omega$  for $T<2J$.
\end{abstract}

\vspace{1cm}


\section { INTRODUCTION }
\label{In}

In this paper we study the finite-temperature properties ($T/J\geq 0.1$) 
of the spin-$\frac{1}{2}$ Heisenberg model coupled to 
(dispersionless) Einstein phonons, 

\begin{eqnarray}
{\cal H} & = & \frac{J}{2}\sum_{l=1}^{N} 
(\vec\sigma_l \vec\sigma_{l+1}-1)(1+g(b_l^{\dag}+b_l)) \nonumber \\
 & & + \Omega\sum_{l=1}^{N} b_l^{\dag}b_l 
\label{H_sp}, 
\end{eqnarray}
in the parameter range 
of the spin-phonon couplings $0 \leq g \leq 1.5$ 
and phonon frequencies $0.2 \leq \Omega/J  \leq 20$.

The phenomenological interest in this model dates back to the 
fundamental work of Pytte \cite{Pytte} 
who showed that a spin-$\frac{1}{2}$ Heisenberg chain 
coupled to three-dimensional 
phonons undergoes a dimerization transition, 
generally referred to as ``spin-Peierls'' transition. 
Motivated by the discovery \cite{Hase} of a transition of 
this type in the inorganic substance \mbox{\cuge,} the model 
proposed by Pytte and its modifications 
have been widely studied in recent time. 
(Thus following a period when \cuge was mostly discussed in the framework 
of a frustrated and dimerized spin-Hamiltonian with a temperature-dependent 
dimerization as a relict of the spin-phonon coupling.) 
In particular the low-temperature properties 
of the model Eq.~(\ref{H_sp}) were evaluated recently by Lanczos and 
DMRG techniques in Refs.\onlinecite{APSA,Augier,Wellein,Weisse,AP} 
and by a continuous time 
world-line Monte Carlo algorithm in Ref. \onlinecite{SC}. 
At $T\neq 0$ a stochastic series expansion 
Monte Carlo  was performed in Ref.\onlinecite{Sandvik}. 

In this paper we are mostly concerned with the finite-temperature properties 
of the model. We thus partly reproduce results of Ref.\onlinecite{Sandvik} 
where the model Eq.~(\ref{H_sp}) was studied 
at finite $T$ for $g=1/(4\sqrt{2})$ and $\Omega=0.2J$. 
In contrast to Ref.\onlinecite{Sandvik} 
we analyze in this work both the adiabatic limit 
of the model and also its gapless region where the thermodynamic 
observables are only slightly effected by the coupling to the phonons. 
We also calculate the specific heat, 
the spin-spin correlations and 
take advantage of the pronounced periodic  patterns 
in the local phonon displacement 
to determine the parameter range where the model is dimerized. 
Our choice of parameters is motivated by Ref.\onlinecite{Werner} 
where the dominating phonon modes for \cuge 
are found to be $\Omega_1 /J=2$, $\Omega_2 /J=4$, $g_1 =-1/8$, and $g_2 =1/2$. 

The outline of the paper is as follows. 
In Sec.~\ref{QM} we briefly describe the Monte Carlo loop 
algorithm we employed. 
In Sec.~\ref{TP} we present the thermodynamic 
properties, in particular we demonstrate 
the presence of a phase transition 
from a gapless to a gapped phase 
by studying the finite-size behavior of 
the magnetic susceptibility. 
We also show the effect of the spin-phonon coupling 
on the specific heat and discuss 
results for the phonon occupation number. 
In Sec.~\ref{DC} we consider the local phonon displacement operator and 
determine the approximate boundary of the dimerized phase of the model. 
We study the dependence of the effective coupling \Jeff on $\Omega$ 
and $g$ and finally find an interesting crossover in the spin-spin 
correlations between the high and low-temperature phase. 
Conclusions are given in Sec.~\ref{Con}. 

\section { QUANTUM MONTE CARLO METHOD }
\label{QM}

For this work we developed an extension of the 
quantum Monte Carlo loop algorithm \cite{ELM,Evertz}. 
Like for the world-line method the key idea of the loop 
algorithm is a mapping of the partition function of a $n$-dimensional 
quantum spin system onto a ($n$+1)-dimensional classical system.  
However, its substantial advantage is that compared to the 
world-line method it allows 
global spin updates, thus substantially reducing the 
autocorrelation times. 
Also the variation of both winding number and 
magnetization are automatically included. 

For our simulation 
we used two Trotter layers for the spin-spin and the spin-phonon 
part of $H$ and introduced 
a third layer for the free phonon part. 
We then combined  global loop updates for the spin-phonon 
and spin-spin coupling  with local heat bath updates modifying the phonon 
occupation number. It is obvious that detailed balance 
is satisfied for the two steps and thus also for the whole procedure. 
The decomposition of the partition function 
which forms the basis of our calculation is given in the appendix.

\begin{figure}[!htb]
\begin{center}
\epsfig{file=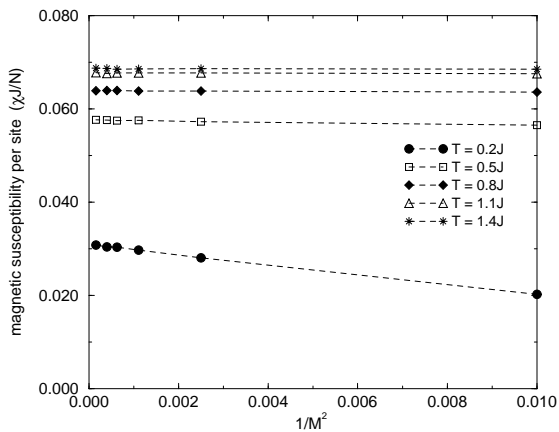,width=8cm}
\end{center}
\caption{ Magnetic susceptibility for 
temperatures $T/J=0.2,0.5,0.8,1.1,1.4$ as a function 
 $1/ (\text{Trotter number})^{2}$ for 
$N=16$ spins, phonon frequency $\Omega =J$, 
and spin-phonon coupling constant $g=0.2$. }
\label{fig1}
\end{figure}

We studied chains with 16 up to 128 sites using 
typically $10^{5} \ldots  10^{6}$ spin updates 
for each temperature. Even though the importance sampling   
technique was employed for the phonon update, 
its acceptance rate is still smaller than that of the spin update. 
Best results were obtained by making $\approx 20$ phonon updates per spin 
update and using only the last of the phonon updates for the 
evaluation of the expectation values. 
Also for each temperature the first 10\% of all the sweeps were used for 
thermalization. We find that the measured 
quantities depend linearly on the inverse Trotter number squared. 
This can be seen in Fig. \ref{fig1}, where we show the susceptibility 
for a $N=16$ system and Trotter number $M=10, \ldots , 80$ as function 
of $1/M^{2}$. In the present work we either extrapolated our 
results in the Trotter number, according to 
the above law, or when we found the fluctuations 
to be larger than the effect of the finite Trotter number, 
we gave the explicit value of the Trotter number in the figures.

\section{THERMODYNAMIC PROPERTIES}
\label{TP}
\subsection{Magnetic Susceptibility}
\label{Su}

It is known that the magnetic susceptibility 
of \cuge in the temperature range above 
the spin-Peierls transition can be well fitted by a frustrated 
Heisenberg model \cite{CCE,RD,us2}. 
It is often objected, however, that 
the excellent agreement of theory and experiment 
is accidental, since it does not 
take into account interchain interactions 
and spin-phonon couplings. 
Motivated by this controversy we pursue the question to what extent  
the spin-phonon coupling influences the susceptibility.

We summarize our observations on the high-temperature 
properties of the susceptibility in the following three points.  

\begin{figure}[!htb]
\begin{center}
\epsfig{file=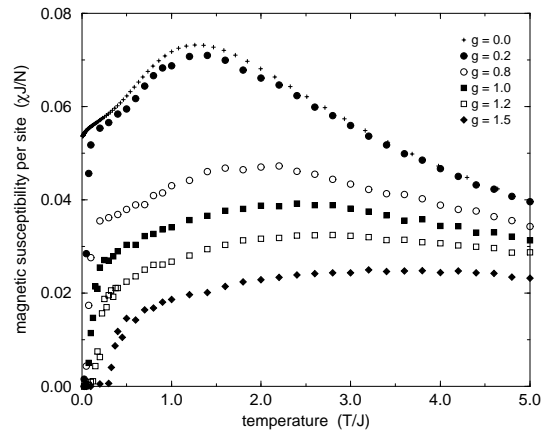,width=8cm}
\end{center}
\caption{Magnetic susceptibility versus temperature 
for  $\Omega =2J$ and 
spin-phonon coupling  $g$ between 0.2 and 1.5. 
The number of sites is $N=64$ for low temperatures, $T\le J$, and 
$N=16$ for high temperatures. 
For comparison we show the exact results for the Heisenberg model 
calculated by A. Kl\"umper \cite{Kluemper} via the quantum transfer 
matrix approach.}
\label{fig2}
\end{figure}

(i) For fixed phonon frequency, the overall height of the susceptibility 
is lowered with an increasing spin-phonon coupling (Fig. \ref{fig2}). 

\begin{figure}[!htb]
\begin{center}
\epsfig{file=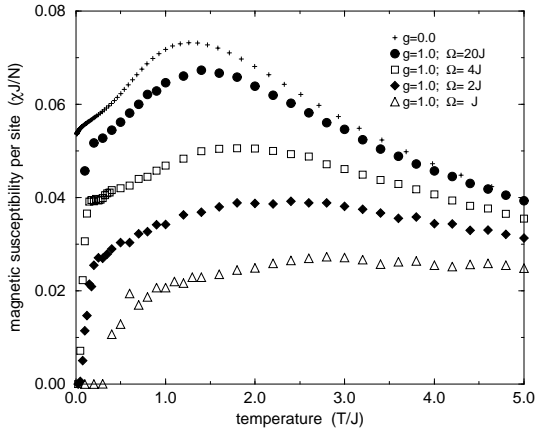,width=8cm}
\end{center}
\caption{
Magnetic susceptibility versus temperature 
for $g=1.0$ and phonon frequency between 
$\Omega =J$ and $\Omega =20J$. The number of sites is $N=64$ for 
low temperatures, $T\le J$ ($T\le 1.5J$ for the $\Omega =J$ chain), and 
$N=16$ for high temperatures. For $g=0$ we show the exact result 
from Ref. \cite{Kluemper}. }
\label{fig3}
\end{figure}

(ii) For fixed $g$, the susceptibility is growing 
and its maximum is shifted to lower temperatures 
with increasing frequency $\Omega$ (Fig. \ref{fig3}). 

(iii) As the spin-phonon coupling is increased 
the maxima of the susceptibility curves are shifted to 
higher temperatures.

Point (i) is not surprising, since an increase of the spin-phonon 
coupling is expected to decrease the magnetic order. Point (ii) 
can be explained by the fact that for 
$\Omega \gg gJ$ the magnetic and phononic degrees of freedom decouple. So for 
$\Omega\rightarrow\infty$ (and $gJ$ fixed) the magnetic properties are 
again dominated by the antiferromagnetic Heisenberg model. 
Point (iii) is of some phenomenological consequence 
because it means that for a substance where the spin-phonon 
coupling cannot be neglected, the value of the spin-spin coupling 
is overestimated, if determined by comparing the experimental data 
with the susceptibility calculated in a pure spin model. 
(We assumed that $J$ is determined by matching temperatures where 
the experimental and theoretical \sus curves have their maxima, which is 
a standard procedure, see for example Ref. \onlinecite{us2}.) 

Also it should be noted that for frustrated models 
$T_{max}/J$ is shifted to lower temperatures with increasing frustration 
parameter, i.~e.~the effect of the spin-phonon coupling is opposite 
to that of frustration. Together with the argument given 
in Section \ref{Ef} it is obvious that the \sus of \cuge 
cannot be fitted with a model of the type  Eq.~(\ref{H_sp}).

To draw definite conclusions from the 
susceptibilities at low $T$, finite-size effects have to be studied carefully. 

\begin{figure}[!htb]
\begin{center}
\epsfig{file=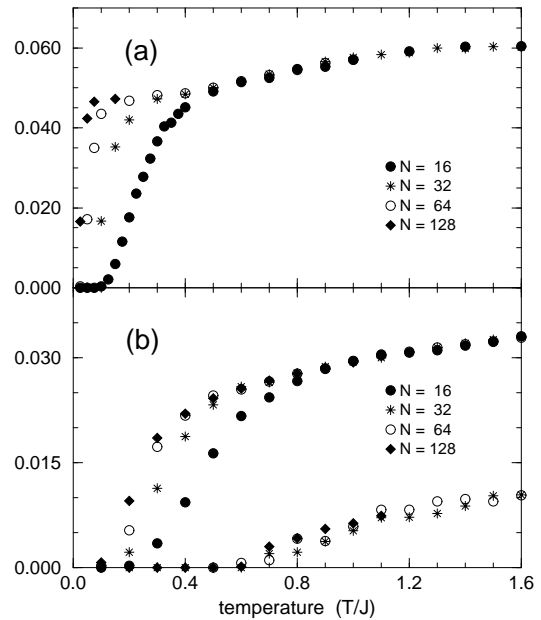,width=14cm}
\end{center}
\caption{
Finite-size behavior of the magnetic susceptibility. 
The upper figure shows a chain with phonon frequency $\Omega =2J$ and 
spin-phonon coupling constant $g=0.5$, the lower figure shows chains 
with $\Omega =J$ and $g=0.8$ (upper curve) and $g=1.5$ (lower curve). }
\label{fig4}
\end{figure}

In Fig. \ref{fig4}a we show results for  $\Omega =2J$ 
and $g=0.5$ and chain lengths $N=16, 32, 64, 128$. 
The tendency of the susceptibility curves is to approach 
a finite value at 
$T\rightarrow 0$. Thus the sharp decrease in the curves for 
$N=16, 32, 64$ is a finite-size effect, i.e. the energy gap causing the drop 
is due to the discreteness of the spectrum of the finite system, it is not 
an intrinsic property of the spin chain (Fig. \ref{fig4}a).

The situation is drastically changed if we consider 
$g=1.5$ and $\Omega=J$ (lower curve in Fig. \ref{fig4}b). 
Here the drop is unchanged when going to larger systems. 
This means that the gap in the spectrum is larger than 
the gap generated by the finite size of the chain. 
The upper curve in Fig. \ref{fig4}b shows results for 
$g=0.8$ and $\Omega=J$, which exhibit clear finite-size effects 
but where the nature of the gap cannot be determined 
from systems of size $N\leq 128$. Since for $g=0$ the model is 
gapless and in the adiabatic limit 
the model is known to be massive, the finite-size behavior 
shown in Fig. \ref{fig4}b gives clear evidence 
that the transition 
between the two phases occurs at finite $g$. 
This transition to a gapped phase is a signature 
of the lattice distortion. Indeed we will 
show in section \ref{Di}, that in the gapped phase we find 
well developed dimerization patterns at low $T$. 

The existence of this phase transition has been previously reported 
at $T=0$ in Ref. \onlinecite{SC} 
where by means of a continuous time world-line algorithm 
it is shown that a critical coupling between gapless and 
spin-fluid phase exists for a fixed $\Omega$. 
Also for a model with a slightly different spin-phonon coupling 
the phase diagram was determined in Ref.\onlinecite {Bursill} 
by means of DMRG technique and for 
a model including a frustrating spin-spin coupling evidence for  
the transition can be found in Ref. \onlinecite{Wellein}. 

\subsection{Specific Heat, Phonon Occupation Number}
\label{Sh}

The idea to get further information about magnetic properties 
by considering besides the susceptibility 
also the specific heat suggests itself 
since the specific heat is an observable, that can be as easily measured and 
calculated as the susceptibility. 
However, in practice 
only for substances with very small $J$ (Cu(L-alanine)$_2$ with a $J=0.55$K 
for example \cite{Souza}) 
the specific heat can 
be directly compared with experiment. 
For most of the inorganic substances (in particular for \cuge) 
which have spin-spin 
couplings of the order of 100K, 
lattice vibrations dominate the specific heat 
and it is an unsolved problem  to extract its magnetic part. 
As a first approximation \cite{Weiden} it was assumed for \cuge 
that it is sufficient 
to subtract the contribution of the phonons 
and consider the rest as the contribution of the spins. 
We show that such an approximation is not 
justified even in the simple case 
of dispersionless phonons. 

\begin{figure}[!htb]
\begin{center}
\epsfig{file=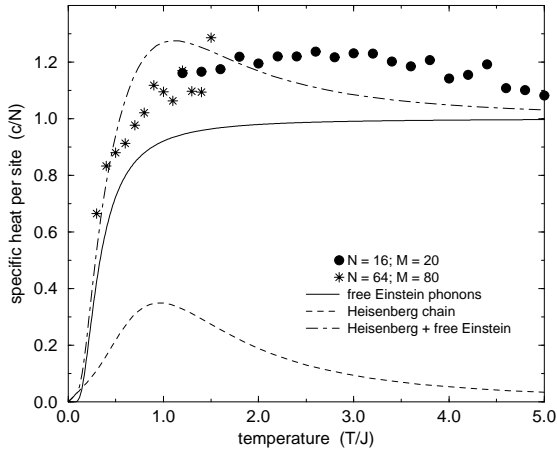,width=8.5cm}
\end{center}
\caption{
Temperature-dependence of the specific heat for a chain 
with phonon frequency 
$\Omega =J$ and spin-phonon coupling constant $g=1.0$. The number 
of sites is 16 and 64, respectively. }
\label{fig5}
\end{figure}

To demonstrate this we plot in Fig. \ref{fig5} 
the specific heat $c$ for a chain with $\Omega =J$ and $g=1.0$ 
together with the specific heat $c_f$  of the free Einstein phonons 
\begin{equation}
c_f = \left( \frac{\Omega}{T} \right)^{2} 
\frac{\text{e}^{\Omega /T}}{ \left( \text{e}^{\Omega /T}-1
\right)^{2}},
\end{equation}
with $\Omega=J$ and the contribution $c_H$ of the isotropic Heisenberg 
model \cite{Kluemper}. 
As can be clearly seen the Einstein phonons are dominant 
at high temperatures, 
in particular the constancy of the specific heat, as expected 
according to the Dulong-Petit rule, is visible. 
However, also the effect of the spin-phonon coupling 
can be perceived, since we find 
that the maximum at $T \approx J$ of the $c_f(T)+c_H(T)$ 
(dot-dashed line) becomes less pronounced and is shifted 
toward higher temperatures by the 
presence of the spin-phonon coupling.  
When going to larger frequencies  (and $g$ fixed), 
the contribution of the 
spin-phonon coupling decreases, but only for  $\Omega\ge 4J$  and $g=1.0$ 
we find that $c(T)$ can be approximated by $c_f(T)+c_H(T)$. 
Of course when considering the  specific heat 
the phonon dispersion, which is not taken into 
account, is most important and more work 
on realistic models is needed to compare theoretical  
results in a quantitative way with experimental data. 

\begin{figure}[!htb]
\begin{center}
\epsfig{file=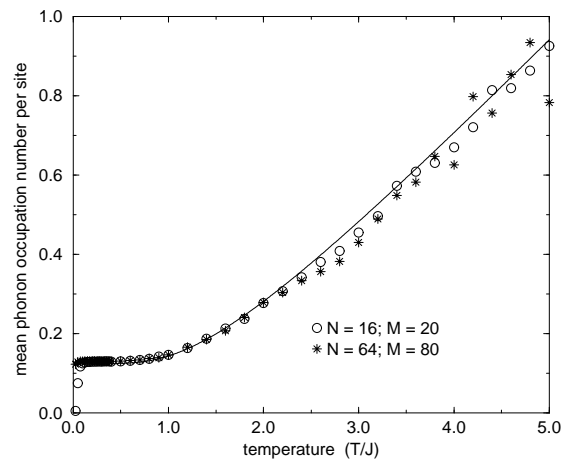,width=8.5cm}
\end{center}
\caption{Mean phonon occupation number 
as a function of temperature for phonon frequency 
$\Omega =4J$ and spin-phonon coupling constant $g=1.0$. 
The number of sites is 16 and 64, respectively. The solid line shows 
$n_f (T)+n_0 (T)$, where $n_0 (T)\equiv n(T)-n_f (T)$ was approximated by the 
temperature-independent value $n_0 (0)=2(gJ/ \Omega )^{2}$.}
\label{fig6}
\end{figure}

Further insight in the model can be gained, by studying to what extent 
the phonon-occupation number is influenced by the spin-phonon coupling.

\begin{equation}
n(T)= \frac{1}{N} \sum_{l=1}^{N} \left< b^{\dag}_l b_l \right> .
\end{equation}
\noindent
We first consider $\Omega=4J$ and $g=1$, parameters for which we found 
above that the specific heat can be approximated without 
the spin-phonon coupling. 
For $n(T)$  we find in this case 
that the effect of the spin-phonon coupling is solely to 
shift the contribution 
\begin{equation}
n_f (T)=  \frac{1}{ \text{e}^{\Omega /T}-1}
\end{equation} 
of the free phonons by a constant value $n_0$ 
(see Fig. \ref{fig6}). 
We determined from the low temperature data that 
$n_0=2(gJ/ \Omega)^{2}$, which in agreement with Ref. \onlinecite{AP} 
where this relation was found for a chain with a single $k=\pi$ 
phonon mode. However, for $g=1$ and $\Omega/J=1,2$ 
deviations from the free phonon case become visible. 
In particular the slope of $n(T)$ at high $T$ is substantially 
smaller than $1/\Omega$. Also the constant shift deviates from 
$2(gJ/ \Omega)^{2}$. 
Note that the drop of $n(T)$ at low $T$ is an effect 
of both finite Trotter number and the finite size of the system.

\begin{figure}[!htb]
\begin{center}
\epsfig{file=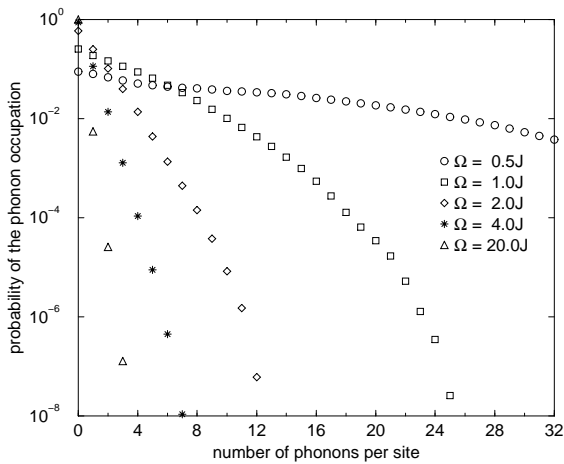,width=8.5cm}
\end{center}
\caption{Probability of the phonon occupation number for 
phonon frequencies  between $\Omega /J=0.5$ and $20$ for a 
chain with $N=64$ sites, 
Trotter number 80, spin-phonon 
coupling constant $g=1.0$, and temperature $T=J$.}
\label{fig7}
\end{figure}

We conclude this section with a more technical remark concerning 
truncation errors. As the dimension 
of the subspaces is quickly growing with the number of 
phonons the limits of a method requiring truncation 
is soon reached, this is  relevant when using 
exact diagonalization which allows to treat only 
$\approx 16$ spins and a few phonons per site. 
At zero temperature the truncation problem can be 
partly overcome  by using coherent phonon states \cite{Augier,Fehren} 
or employing the DMRG technique, which allows to treat system sizes 
of up to 256 spins\cite{Bursill}. 
At finite temperatures, however, the truncation problem is even more 
important. With the probability distribution 
of the phonon occupation number 
shown in Fig. \ref{fig7} 
we dramatically demonstrate that high cutoffs are essential 
when studying finite temperature properties. 
For the correct treatment  of a chain with 
$\Omega \ge J$ and $g=1.0$  at $T=J$ 
a truncation at $\approx 30$ spins is needed (as can be seen from 
the figure the probability for more than 28 phonons per site 
is less than $10^{-7}$). 
For $\Omega/J=0.5$ the truncation is  $\leq 50$ spins/site 
and only very high phonon frequencies as $\Omega/J= 20$ can 
be treated with 4 phonons per site.

\section{ Dimerization and Spin-Correlations }
\label{DC}

Addressing now the question of the dimerization, we 
expect that the ground state of 
the model Eq.~(\ref{H_sp}) 
in the massive phase 
shows short range correlations and strong signatures of dimerization. 
We extract the boundaries of the phase 
diagram in the $g$-$\Omega$ plane by studying the 
dimerization patterns at low temperatures 
and corroborate the results by considering the divergence of 
the structure factor at $q=\pi$. 
We also give account of the averaged phonon displacement 
often referred to as an ``effective coupling'', 
which is nonzero for all $g\neq0$ and $\Omega\neq0$.

\subsection{ Dimerization }
\label{Di}

To begin, let us recall the definition of the 
local phonon displacement operator 
\begin{equation}
\delta_l = g \left< b^{\dag}_l + b_l \right> 
\end{equation}
which  corresponds in the adiabatic limit 
to the dimerization in a Heisenberg model with alternating couplings. 
Its  average over all sites $l$, 
\begin{equation}
\delta = \frac{1}{N} \sum_{l=1}^{N} \delta_l ,
\end{equation}
represents the uniform phonon displacement 
and is closely related 
to the effective spin-spin coupling \cite{Pytte,Sandvik} 
\begin{equation}
J_{\text{eff}} =J(1+ \delta ).
\label{Jeff} 
\end{equation}

It is well-known that in spin-Peierls substances like \cuge 
the transition from the uniform to the dimerized phase 
occurs at finite temperatures, here however, we consider a strictly 
one-dimensional model with a 
dimerization transition at $T=0$. 
So to encounter dimerization one needs to go to temperatures 
as low as possible.

\begin{figure}[!htb]
\begin{center}
\epsfig{file=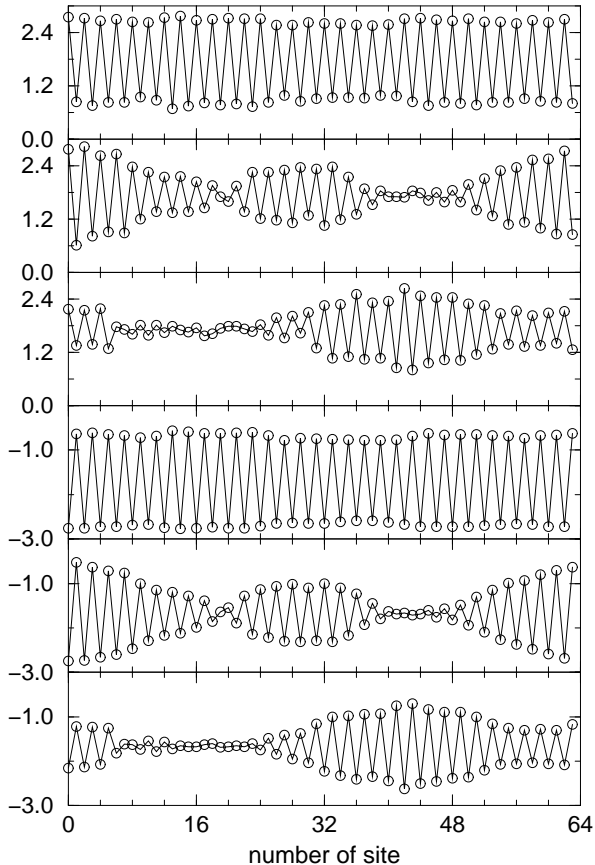,width=16cm}
\end{center}
\caption{ 
Upper three graphs:  local displacement 
$\left< b_{l}^{\dag}+b_l \right> $ for 
$T=0.3J, 0.4J, 0.5J$. 
Lower three graphs: spin-spin-correlation 
$\left< \vec\sigma_l \cdot\vec\sigma_{l+1} \right> $ 
for the same temperatures. 
All graphs are for $N=64$, $M=80$, $\Omega =2J$, 
$g=1.5$, and averaged over 10,000 spin updates. }
\label{fig8}
\end{figure}

By studying $\delta_l$ at low temperatures in the parameter-range 
$0 \leq \Omega /J \leq 3 $ and $0 \leq g \leq 2 $ 
we can clearly distinguish two different phases 
of the model.  We find either pronounced periodic patterns in $\delta_l$ 
(see  Fig. \ref{fig8}) or 
random fluctuations with no perceivable periodicity 
down to temperature $T/J=0.05$. 
We note that as $T$ is risen to higher temperatures the periodic 
patterns weaken and at higher $T$ 
domains of periodic fluctuation are perceivable, 
which reflect a 
``reminiscence'' of the periodic fluctuations present at $T=0$ 
(see second and third graph of Fig. \ref{fig8}).

It is of interest to see how the periodic 
patterns in the phonon displacement are connected with the 
formation of spin-dimers. So to 
complete the picture  we calculated 
the local spin-spin correlations 
$\left< \vec\sigma_l \cdot\vec\sigma_{l+1} \right> $. 
We find that 
the periodic patterns in the local displacement $\delta_l $ 
are well in phase with the formation of spin singlet states, 
which demonstrates the 
coupling of the phononic degrees of 
freedom  and the spin degrees of freedom. 
This is also an \'a posteriori check of 
the consistency of our combined spin and phonon updates. 
(see the lowest three graphs of Fig. \ref{fig8}. As can be seen 
from the Fig. \ref{fig8} the dimerization is not complete, 
since the correlation between the singlet states is not zero.)

Thus $\delta_l$ proves to be a valuable tool, which allows 
more easily than the susceptibility to decide, for fixed parameter 
values whether the model is in a dimerized state. 

\begin{figure}[!htb]
\begin{center}
\epsfig{file=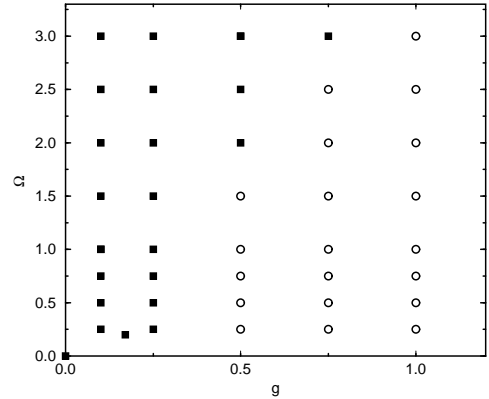,width=7.5cm}
\end{center}
\caption{ 
Phase-diagram of the model in the $g$-$\Omega$ plain. 
Circles mark points with dimerization, squares points where 
no dimerization was found for $N=64$ and $T\geq0.1J$.}
\label{fig9}
\end{figure}

We use this  property 
to determine the approximate phase boundaries in the $g-\Omega$ 
plain. To be concrete, we calculated 
$\delta_l$ for a 64 site system
at temperatures $T/J=0.4,\dots,0.1$ and inferred from 
the periodic fluctuations the 
phase diagram shown in Fig. \ref{fig9}. 
Since for all points $\delta_l$ either shows random patterns 
or the periodic fluctuations, 
we can unambiguously determine, 
whether the model at a certain point in 
parameter space is in a dimerized phase. 
We cannot however state 
definitely, that the model is gapless 
if we encounter no dimerization, because 
we are limited both in system size and 
temperature. So rigorously 
the separation line between the two phases in Fig. \ref{fig9} 
gives the minimal extension
of the dimerized phase, i.e. 
in the thermodynamic limit the dimerized phase will be larger and 
the separation line in Fig. \ref{fig9} will be shifted to the left.
Comparing with the results of Fig. \ref{fig4}   
we find that 
the point $g=1.5$ and $\Omega = J$ 
where from the susceptibility we found that the model must be in a massive 
phase shows clear patterns of dimerization, and the point 
$g=0.8$ and $\Omega = J$, for which 
the finite size argument was not conclusive, turns out to be 
in the dimerized phase. 

\subsection{ Effective coupling \Jeff }
\label{Ef}

\begin{figure}[!htb]
\begin{center}
\epsfig{file=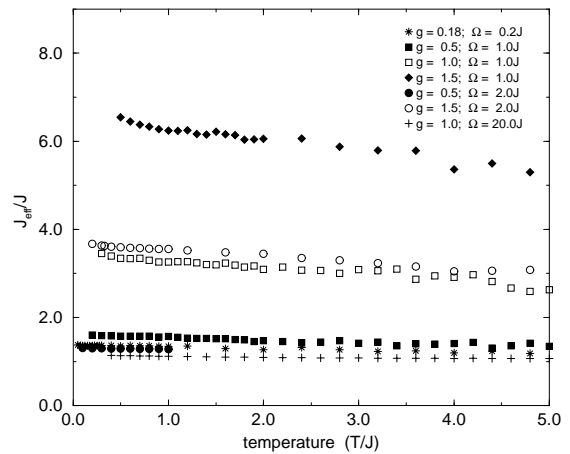,width=8.5cm}
\end{center}
\caption{
Effective spin-coupling $J_{\text{eff}}/J$ versus temperature.} 
\label{fig10}
\end{figure}

We turn now to the averaged quantity $\delta$ 
and study the effective coupling \Jeff as given 
by Eq.~(\ref{Jeff}). It is worth noting that both in the dimerized 
and the undimerized phase $\delta$ is nonzero i.e. 
the spin-spin coupling is always shifted by the presence 
of the spin-phonon coupling. 
Our analysis shows that for all choices of parameters 
$J_{\text{eff}}$ is a slowly decreasing function of temperature 
(Fig. \ref{fig10}), this behavior has been noted before 
for $g=1/(4\sqrt{2})$ and $\Omega=0.2J$ in 
Ref. \onlinecite{Sandvik}. 

>From the phenomenological point it is interesting 
that a lowering of $J$ with temperature is in agreement with 
Ref. \onlinecite{FL} where 
a comparison of inelastic neutron scattering 
data on \cuge with calculated dynamic structure factors $S(q,\Omega)$ 
gave evidence for a decrease of $J$ with increasing temperature. 
A similar behavior of $J$ 
has been reported in Ref. \onlinecite{Uhrig1}.

\begin{figure}[!htb]
\begin{center}
\epsfig{file=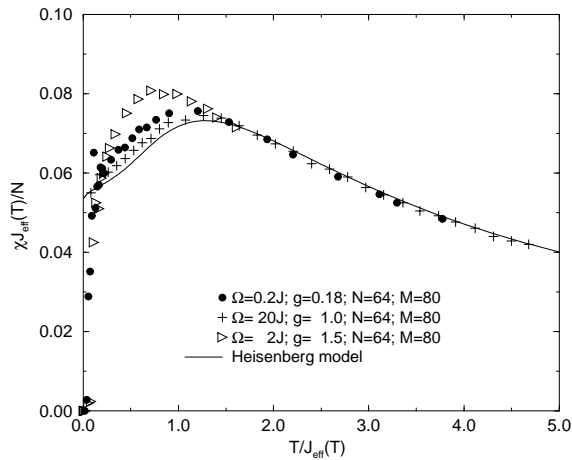,width=8.5cm}
\end{center}
\caption{ Rescaled susceptibilities 
$\chi (J,g,T, \Omega ) J_{\text{eff}}(T)/N$ 
versus $T/J_{\text{eff}}(T)$}
\label{fig11}
\end{figure}

With our results for \Jeff we are in the position, 
to study the question to what extent 
$J_{\text{eff}}$ is an effective coupling in the proper 
sense, in other words, whether it can be used 
to rescale the thermodynamic quantities in such a way that 
the results for the Heisenberg model are regained. 
We find that the rescaled susceptibilities $\chi(T)J_{\text{eff}}(T)$ 
plotted versus temperature as shown in 
Fig. \ref{fig11} 
do not fall together with the curves of the Heisenberg model. 
So one is forced to conclude that the effective coupling 
cannot be used to bring the curves with different parameter together. 
One  cannot infer from this however, 
that  essential new physics 
-- maybe a sign of a frustration -- 
is perceivable. The above result  means only that \Jeff 
is not the correct variable to rescale the thermodynamical 
observables. 

\begin{figure}[!htb]
\begin{center}
\epsfig{file=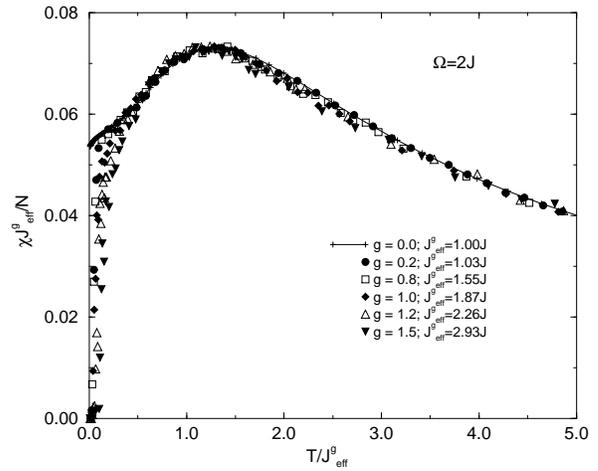,width=8.7cm}
\end{center}
\caption{ Rescaled susceptibilities 
$\chi (J,g,T, \Omega ) J_{\text{eff}}^g /N$ 
versus $T/J_{\text{eff}}^g$}
\label{fig12}
\end{figure}

On the other hand we find that to high precision the susceptibilities 
(for not too low $T$) can be rescaled by a global and temperature 
independent $J_{\text{eff}}^g /J$ (Fig.~\ref{fig12}). 
That is we find 
\begin{equation}
\chi(J,g,T,\Omega)J_{\text{eff}}^g =
J\chi^{\text{AFH}}(T/J_{\text{eff}}^g,J_{\text{eff}}^g)
\end{equation}
where $J_{\text{eff}}^g$ is a global effective 
coupling , which we determined by comparing the temperatures 
$T_{max}$ 
where the susceptibilities of the Heisenberg model 
and the model Eq.~(\ref{H_sp}) have their maxima 
\begin{equation}
J_{\text{eff}}^g=J \frac{T_{\text{max}}(g,\Omega)}
{T_{\text{max}}^{\text{AFH}}}.
\end{equation}
We note that there is a discrepancy to Ref. \onlinecite{Sandvik} 
where an overall $g$-Factor was required 
to obtain a similar scaling law. 

This scaling of the susceptibilities is of phenomenological importance also 
for \cuge, since it is well known that the susceptibility 
of a frustrated model \cite{CCE,us2} cannot be mapped by a rescaling of $J$ 
onto that of the Heisenberg model. This lends further evidence to the argument 
given in Sec. \ref{Su} that no agreement with the experimental \sus 
of \cuge can be achieved. 
By checking with 
data from the model with frustration we estimate that a 
frustration parameter of $\approx 0.12$ 
might go undetected in our analysis, because 
the statistical fluctuations hide the small 
effect on $\chi$ for frustrations 
less than 0.12. 

\begin{figure}[!htb]
\begin{center}
\epsfig{file=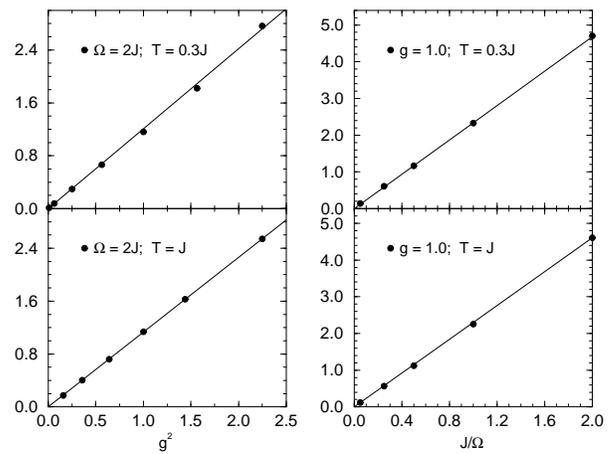,width=8.5cm}
\end{center}
\caption{
Displacement $\delta = J_{\text{eff}}/J-1$ versus $g^{2}$ (left side) 
and versus $J/ \Omega$ (right side) for $N=64$ and Trotter number $M=80$. }
\label{fig13}
\end{figure}

As a last point we show that  $\delta$ 
is proportional to $ g^{2}/ \Omega $ 
in the temperature range $T<2.0J$, 
(see Fig.  \ref{fig13}. 
The constants of proportionality for the data shown in 
Fig.~\ref{fig13} are 
$c(T=0.3J)=(2.38\pm 0.04)J$ and $c(T=J)=(2.28\pm 0.02)J$.) 
At higher temperatures we found deviations from the simple law. 
This result is in qualitative agreement with observations 
by Wellein et al. \cite{Wellein}, who found a lowering of 
the dimerization with growing frequency at $T=0$, 
it also confirms calculations of Pytte \cite{Pytte} 
and  Uhrig\cite{Uhrig2}, who 
found the spin-phonon Hamiltonian (for translationally 
invariant phonons) equivalent 
to an effective spin Hamiltonian with spin-spin coupling 
$\propto g^{2}/ \Omega $. 

\subsection{ Correlations }
\label{Co}

With our method it is straightforward to calculate 
spin-spin-correlation functions \mbox{$<\vec S_i \cdot\vec S_j >$} 
as a function of temperature. 
We postpone a 
fit of the correlation functions 
in the strongly dimerized phase and a detailed discussion of
the critical exponents in the gapless phase to 
a later publication 
and highlight at the end of this work 
the most spectacular features of the correlation functions only.

\begin{figure}[!htb]
\begin{center}
\epsfig{file=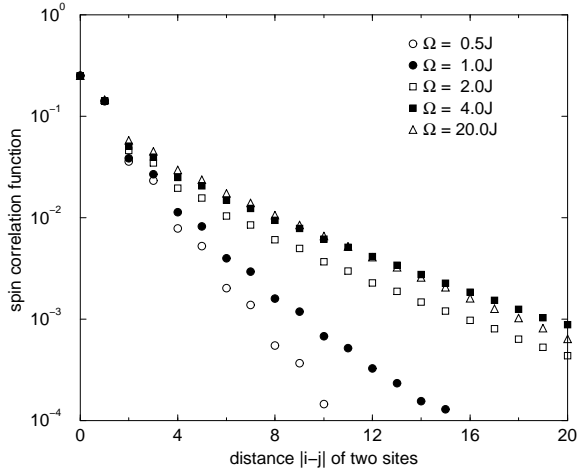,width=8.7cm}
\end{center}
\caption{
$<\vec S_i \cdot\vec S_j > (-1)^{i-j}$, 
as a function of distance 
for $g=1.0$ for a 64 site chain, 
M=80, phonon frequency $\Omega =0.5J,\ldots,20J$ at $T=0.3J$. }
\label{fig14}
\end{figure}

(i) We find that the  spin-spin correlations as a function of distance 
show a pronounced  steplike behavior (see Fig. \ref{fig14}) 
in the gapped phase. 
This is  again a signature of 
the  dimerization of the ground state. 
(Note that for a completely dimerized 
state we would have one step only, however, 
we find a finite correlation between the dimers.)

(ii) For increasing $\Omega$ at fixed $g$ one 
finds the correlations more and more unaffected by the spin 
phonon coupling, which is due to 
the decoupling of spins and phonons as we observed before 
when studying the susceptibility.

\begin{figure}[!htb]
\begin{center}
\epsfig{file=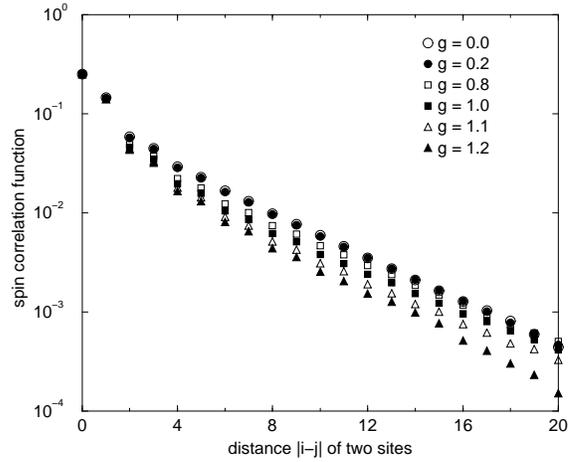,width=8.5cm}
\end{center}
\caption{Spin-spin correlation 
$<\vec S_i \cdot\vec S_j > (-1)^{i-j}$, 
as a function of distance 
for $g=0,\ldots ,1.2$ for a 64 site chain, 
M=80, phonon frequency $\Omega =2J$ at $T=0.3J$.}
\label{fig15}
\end{figure}

(iii) As a function of $g$ we observe a
crossover between the high-temperature 
and the low-temperature range. 
At low temperatures the correlations 
decrease with growing spin-phonon coupling (Fig. \ref{fig15}). 
This is analogous to the dimerized model where one finds a 
lowering of the correlation with growing dimerization. 
For high temperatures, however, the inverse behavior 
occurs: 
the correlations for $g=1.5$ are highest and for $g=0.2$ are lowest. 
(For $\Omega =2J$ the transition between low and high-temperature region 
occurs at $T=(0.55 \pm 0.05)J$.)

\begin{figure}[!htb]
\begin{center}
\epsfig{file=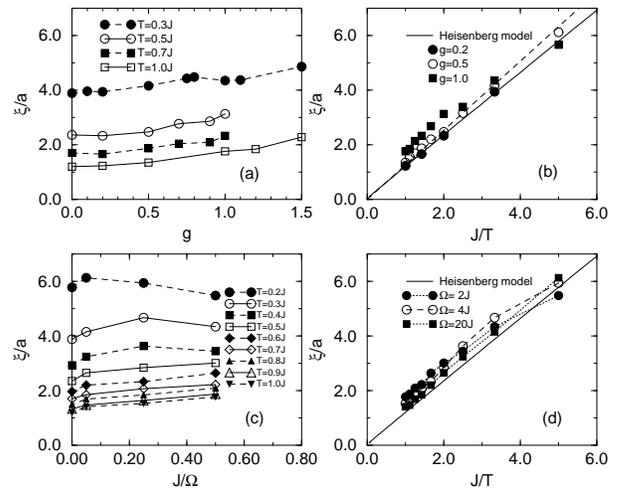,width=8.2cm}
\end{center}
\caption{ Correlation length $\xi$ in lattice units $a$ for $\Omega =2J$ 
(upper figures) and $g=1.0$ (lower figures). The chain length is $N=64$ 
and the Trotter number $M=80$.}
\label{fig16}
\end{figure}

(iv)  In the gapless region of the model the correlations 
decay  exponentially 
with a correlation length that coincides 
for $g\leq0.2 $ with the known Heisenberg result 
\begin{equation}
\xi(T) =c T^{- \nu} \ \ \text{with} \ \ \nu=1 \ \ 
\text{for}\ \  N\rightarrow\infty
\label{corr} 
\end{equation}
(see Fig.~\ref{fig16}b). 
For larger $g$ we encounter deviations from 
Eq.~(\ref{corr}) 
which can best be fitted by 
allowing $\nu$ to differ from 1. 
(e.~g. 
$\nu =0.92$ for $g=0.5$ and $\Omega =2J$). 
In Fig.~\ref{fig16}a,c 
we show  $\xi$ 
as a function of $g$ and $1\over \Omega$. 
We find that  for 
 temperatures $0.5<T/J<1 $ the correlation length  
$\xi$ decreases slightly with $g$ 
and $1\over \Omega$. 
However, a more careful investigation is needed to 
give conclusive evidence for the functional 
dependence of $\xi(g,\Omega ,T)=c(g,\Omega )T^ {- \nu(g,\Omega)}$. 

\begin{figure}[!htb]
\begin{center}
\epsfig{file=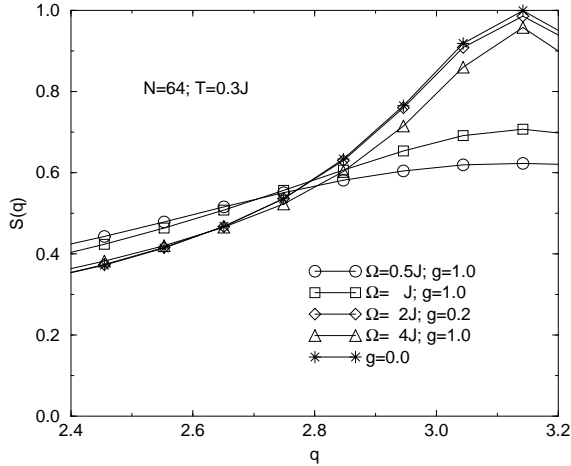,width=8.7cm}
\end{center}
\caption{ Structure factor $S(q)$ versus momentum $q$.}
\label{fig17}
\end{figure}

(v) The pronounced maxima of the 
static structure factor 
$S(q,T)$ at $q=\pi$ for $g=0$ as shown in 
Fig.~\ref{fig17} 
originate from  
the logarithmic divergence 
of $S(q,T=0)$ of the  isotropic Heisenberg model. 
(For a study as to what extent this divergence 
can be seen at finite $T$ we refer to Ref. \onlinecite{FML}.) 
For the model 
Eq.~(\ref{H_sp}) in the gapless phase the peak in $S(q)$ at $q=\pi$ 
is only slightly diminished as compared to the 
Heisenberg model, indicating that the models 
show a similar divergent behavior. 
On the contrary in the gapped phase the 
structure factor shows a broad maximum hinting toward a regular behavior 
of $S(q=\pi,T=0)$. 

\section{Conclusions}
\label{Con}
We have developed a modified quantum Monte Carlo loop algorithm to study 
the finite-temperature properties of the one-dimensional isotropic 
antiferromagnetic Heisenberg model coupled to Einstein phonons. 
We have investigated in detail the magnetic susceptibility, specific heat, 
phonon occupation number, dimerization, and spin-correlation  
for chains with 16 up to 128 sites down to temperatures $T=0.05J$.

We found evidence for a phase transition from a gapless to a massive phase 
and used the periodic fluctuations in the local displacement 
to determine the approximative phase diagram of the model. 
Furthermore, the dimerization patterns we encountered were accompanied by 
the formation of spin singlet states.

The temperature-dependence of 
the effective spin-spin coupling $J_{\text{eff}}$ was calculated 
and it was shown that 
$J_{\text{eff}}(T)=(1+c(T)g^{2}/ \Omega )J$, 
for $T/J \leq 2.0$. 

At low temperatures the mean phonon occupation number was found to be 
proportional to the square of the spin-phonon coupling constant and inversely 
proportional to the square of the phonon frequency. 

We demonstrated that the susceptibility of \cuge cannot 
be described by the model.

\section{Acknowledgement}
We thank K.~Fabricius for his continuous and stimulating interest 
in our work. We are grateful to A.~Fledderjohann, 
G.~Uhrig, W.~Weber, and R.~Werner for valuable discussions. 
The support of the Deutsche 
Forschungsgemeinschaft and the Graduiertenkolleg Wuppertal is 
acknowledged.

\section{Appendix}
\label{Ap}

The Trotter decomposition of the partition 
function $Z$ we used as basis of our quantum Monte Carlo method reads:

\begin{eqnarray}
Z & = & \lim_{M\rightarrow\infty} \sum_{k_0 ,\ldots ,k_{3M-1}} 
        \prod_{j=0}^{M-1} \nonumber \\
 & & \left( \prod_{l=0}^{N-1} 
        <k_{3j,l}| \text{e}^{-\beta\Omega (b_{l}^{\dag}b_l)/M} |
        k_{3j+1,l}> \right)  \nonumber \\
 & & ( \prod_{l=0}^{N/2-1} <k_{3j+1,2l}| 
        \text{e}^{- \beta {\cal H}_{2l}/M} | k_{3j+2,2l}>  \nonumber \\ 
 & &  <k_{3j+2,2l+1}| \text{e}^{- \beta {\cal H}_{2l+1}/M}|k_{3j+3,2l+1}> ) .
\end{eqnarray}
$k$ denotes the spin and phonon configurations of the lower and upper 
edge of the interacting plaquettes, the inverse temperature is 
$\beta$, the Trotter number $M$, and the Hamiltonian for the 
spin-spin and the spin-phonon interaction is 
\begin{equation}
{\cal H}_l = \frac{J}{2} (\vec\sigma_l \vec\sigma_{l+1}-1)
(1+g(b_l^{\dag}+b_l)) .
\end{equation}

\end{document}